\documentclass[12pt,preprint]{aastex}

\usepackage{epsfig}
\usepackage{lscape}

        \shorttitle{O-Star Velocities}
        \shortauthors{Williams et al.}

\begin{document}


\title{Radial Velocities of Galactic O-Type Stars. II. Single-lined
  Spectroscopic Binaries}

\author{S. J. Williams, D. R. Gies}
\affil{Center for High Angular Resolution Astronomy and \\
 Department of Physics and Astronomy,\\
 Georgia State University, P. O. Box 4106, Atlanta, GA 30302-4106; \\
 swilliams@chara.gsu.edu, gies@chara.gsu.edu}

\author{T. C. Hillwig\altaffilmark{1}}
\affil{Department of Physics and Astronomy, Valparaiso University, Valparaiso,
  IN, 46383; todd.hillwig@valpo.edu}

\author{M. V. McSwain\altaffilmark{1}}
\affil{Department of Physics, Lehigh University, 16 Memorial Drive East,
  Bethlehem, PA 18015; mcswain@lehigh.edu}

\author{W. Huang\altaffilmark{2}}
\affil{Department of Astronomy, University of Washington,
Box 351580, Seattle, WA 98195-1580; hwenjin@astro.washington.edu }

\altaffiltext{1}{Visiting Astronomer, Cerro Tololo Inter-American Observatory, 
  National Optical Astronomy Observatory, which are operated by the
  Association of Universities for Research in Astronomy, under contract with 
  the National Science Foundation.}
\altaffiltext{2}{Visiting Astronomer, Kitt Peak National Observatory,
  National Optical Astronomy Observatory, which is operated by the Association
  of Universities for Research in Astronomy (AURA) under cooperative agreement with
  the National Science Foundation.}

\begin{abstract}

We report on new radial velocity measurements of massive stars
that are either suspected binaries or lacking prior observations.
This is part of a survey to identify and characterize
spectroscopic binaries among O-type stars with the goal of
comparing the binary fraction of field and runaway stars with
those in clusters and associations.
We present orbits for HDE 308813, HD 152147, HD 164536, 
BD$-16^\circ 4826$ and
HDE 229232, Galactic O-type stars exhibiting single-lined 
spectroscopic variation. By fitting model spectra to our 
observed spectra we obtain estimates for effective temperature, 
surface gravity, and rotational velocity. We compute orbital periods and
velocity semiamplitudes for each system and note the lack
of photometric variation for any system. These binaries probably appear
single-lined because the companions are faint and because their
orbital Doppler shifts are small compared to the width of the
rotationally broadened lines of the primary.

\end{abstract}


\setcounter{footnote}{2}

\section{Introduction}

The multiplicity frequency of O-type stars in clusters or
associations is large. \citet{san11} find that $\approx 44\%$
of O-stars in open clusters are members of relatively short period,
spectroscopic binairies. \citet{mas98,mas09} considered the total 
numbers of spectroscopic (short period) and angularly resolved
binaries (long period) and derived an estimate of $\approx 75\%$
for O-stars with some companion, but this estimate must be a
lower limit because the intermediate period range (years to
decades) remains largely unexplored.
Spectroscopic detection of companions to O stars
presents many problems. The high luminosities of early-type
stars can easily outshine spectroscopic companions with a difference 
of just a few temperature subtypes, and/or a single 
luminosity class. Some O stars are rapid rotators with
broad lines that can hide the spectral features of a companion,
even with high resolution spectra. Systems may also be seen at
an unfavorable inclination to our line of sight, reducing the 
shifts of spectral features that would be observed. Some 
combination of these three scenarios results in the detection
of such systems as low-amplitude, single-lined
spectroscopic binaries (SB1), or no detection of velocity
variability at all.

We present results on five such systems here that were observed
as part of a larger group of O-type stars with little
prior observational work and that
appear with a binary designation of ``U'' (unknown) or ``?'' 
(uncertain) in \citet{mas98}. 
The current status of each of these systems is either 
unknown spectroscopic variability (HDE 229232 and BD$-16^\circ 4826$) 
or known variability but with insufficient data to compute an orbit
(HDE 308813, HD152147, and HD 164536). With new data presented
here (\S 2), we measure radial velocities, determine orbital elements,
and compute stellar parameters for each system (\S 3), closely 
following the procedure used to analyze velocity non-variables 
in Paper I of this series \citep{wil11}. 
We discuss these first single-lined spectroscopic 
orbits and physical parameters for each system (\S 4). 
We conclude the paper with a discussion about the nature of
the unseen companions (\S 5).


\section{Observations}

Observations of HDE~308813, HD~152147, BD$-16^\circ 4826$, and HD~164536 
were obtained from 
2003 March 16 to 20 UT, 2004 January 13 to 29 UT, and May 27 to June 8 UT
with the Cerro Tololo Inter-American Observatory (CTIO) 1.5 m telescope 
and RC spectrograph. We used grating $\#47$ (831 lines per mm, 
$8000$~\AA~blaze wavelength in the first order Littrow configuration) in
second order together with a BG39 or CuSO$_4$ order blocking filter 
with a slit width of $2''$. The detector was a
Loral $1200\times 800$ CCD. This arrangement produced spectra
covering the range 4058 -- 4732 \AA ~with a FWHM resolving power of
$R=\lambda / \Delta \lambda \approx 2750$ and a wavelength calibration
accuracy of $\sim$4 km~s$^{-1}$ based on the rms scatter of fits
to comparison spectra. The instrumental broadening from this setup, 
found from the FWHM of the comparison lines, was computed to be
equivalent to a Doppler broadening of 109 km~s$^{-1}$. 
Exposure times were several
minutes in order to reach a signal-to-noise ratio (S/N) exceeding 100 
per pixel. Each observation was bracketed
with HeAr comparison spectra for wavelength calibration, and
numerous bias and flat field spectra were obtained each night.

HDE~229232 was observed with the Kitt Peak National
Observatory (KPNO) 2.1 m telescope during an observing run from 2008 
November 15 to 21 UT.
These observations made use of the Goldcam spectrograph with
grating G47 (831 grooves mm$^{-1}$) in second order with a CuSO$_4$
order sorting filter.  The detector was the T3KC CCD
(a $3072\times1024$ pixel array with $15\times 15$ $\mu$m pixels),
and the slit width was set to $1\farcs3$ resulting in spectra with a 
FWHM resolving power of $R = \lambda / \Delta \lambda = 3030$.  
The exposure time for HDE 229232 was 600 s, enabling a S/N of
about 200 per pixel in the continuum.
The wavelength range is 3942 -- 5032 \AA ~with a
wavelength calibration accuracy of $\sim$5 km~s$^{-1}$ based on the rms
scatter of fits to comparison lines. Instrumental broadening for these data,
again estimated from the FWHM of comparison lines, 
was found to be 99 km~s$^{-1}$.

All spectra were reduced using standard routines in IRAF\footnote{IRAF 
is distributed by the National Optical Astronomy Observatory, 
which is operated by the Association of Universities for Research 
in Astronomy, Inc., under cooperative agreement with the National 
Science Foundation.} and then rectified to unit continuum via fitting 
line-free regions. Next, all spectra for a specific target were transformed 
to a common heliocentric wavelength grid in $\log\lambda$ increments and
were placed into a two dimensional array as a function of spectrum 
wavelength and time. The subsequent velocity measurements and analysis
of spectra may then be done on the entire array of spectra, 
rather than on each individual spectrum. 


\section{Radial Velocities and Stellar Parameters}\label{sectrv}

To measure both velocities and stellar parameters, we began
with an approximate model template for the star. A simple
``by eye'' fit to several prominent absorption features 
was made by interpolating in the OSTAR2002 grid \citep{lan03} for
stars with $T_{\rm eff}>30$ kK and in the BSTAR2006 grid \citep{lan07}
for stars cooler than 30 kK.

Velocities for each spectrum for each object were then 
measured using the standard method of cross-correlation, and the
uncertainty in the resulting velocity was estimated using the
method of \citet{zuc03}.
The model was cross-correlated with each spectrum in 
the particular object's spectrum stack. Prior to 
cross-correlation, object dependent emission 
features and interstellar features were removed, including the
broad diffuse interstellar band (DIB) near $\lambda$4428 \AA.
The radial velocities are listed in Table~\ref{sb1rvs} 
with targets ordered by right ascension. The table lists 
the object name, Heliocentric Julian date of observation, 
orbital phase (see below), radial velocity, velocity 
uncertainty, and the observed minus calculated $(O-C)$ 
value for each measurement. Note that the quoted 
velocity uncertainties do not include those uncertainties
related to the wavelength calibration of the spectra.

\begin{figure}
\begin{center}
{\includegraphics[angle=90,height=12cm]{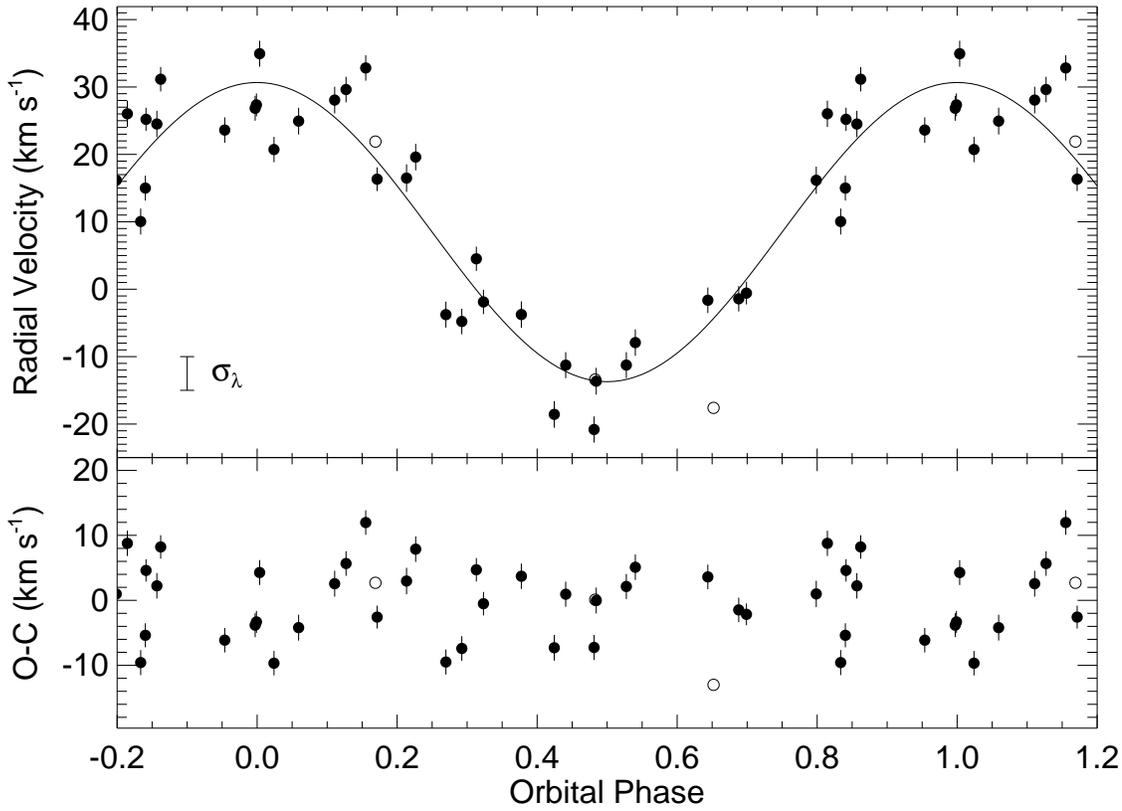}}
\end{center}
\caption{Radial velocities for HDE 308813. Filled circles are data
  from this analysis with uncertainties shown by line segments,
  and open circles are data from \citet{hua06}. Plotted beneath the 
  velocities are the observed minus calculated values based on 
  the circular orbital fit. Also shown is the 
  5 km s$^{-1}$ uncertainty associated with the wavelength calibration.}
\label{hd308813fit}
\end{figure}

\begin{figure}
\begin{center}
{\includegraphics[angle=90,height=12cm]{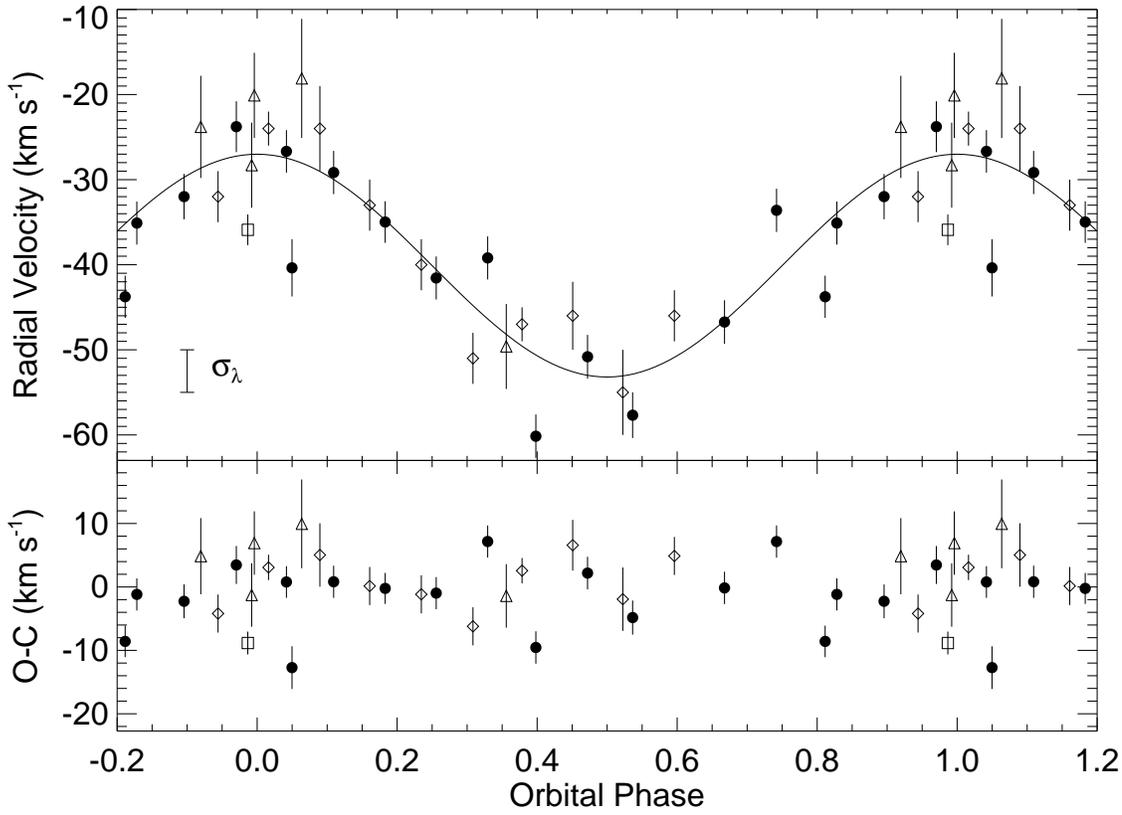}}
\end{center}
\caption{Radial velocities for HD 152147. Filled circles are data
  from this analysis, the open square is from \citet{con77a}, 
  open triangles are from \citet{str44}, and open diamonds are
  from \citet{lev88}, with uncertainties shown by line segments. 
  Plotted beneath the velocities are the observed minus calculated 
  values based on the circular orbital fit.}
\label{hd152147fit}
\end{figure}

\begin{figure}
\begin{center}
{\includegraphics[angle=90,height=12cm]{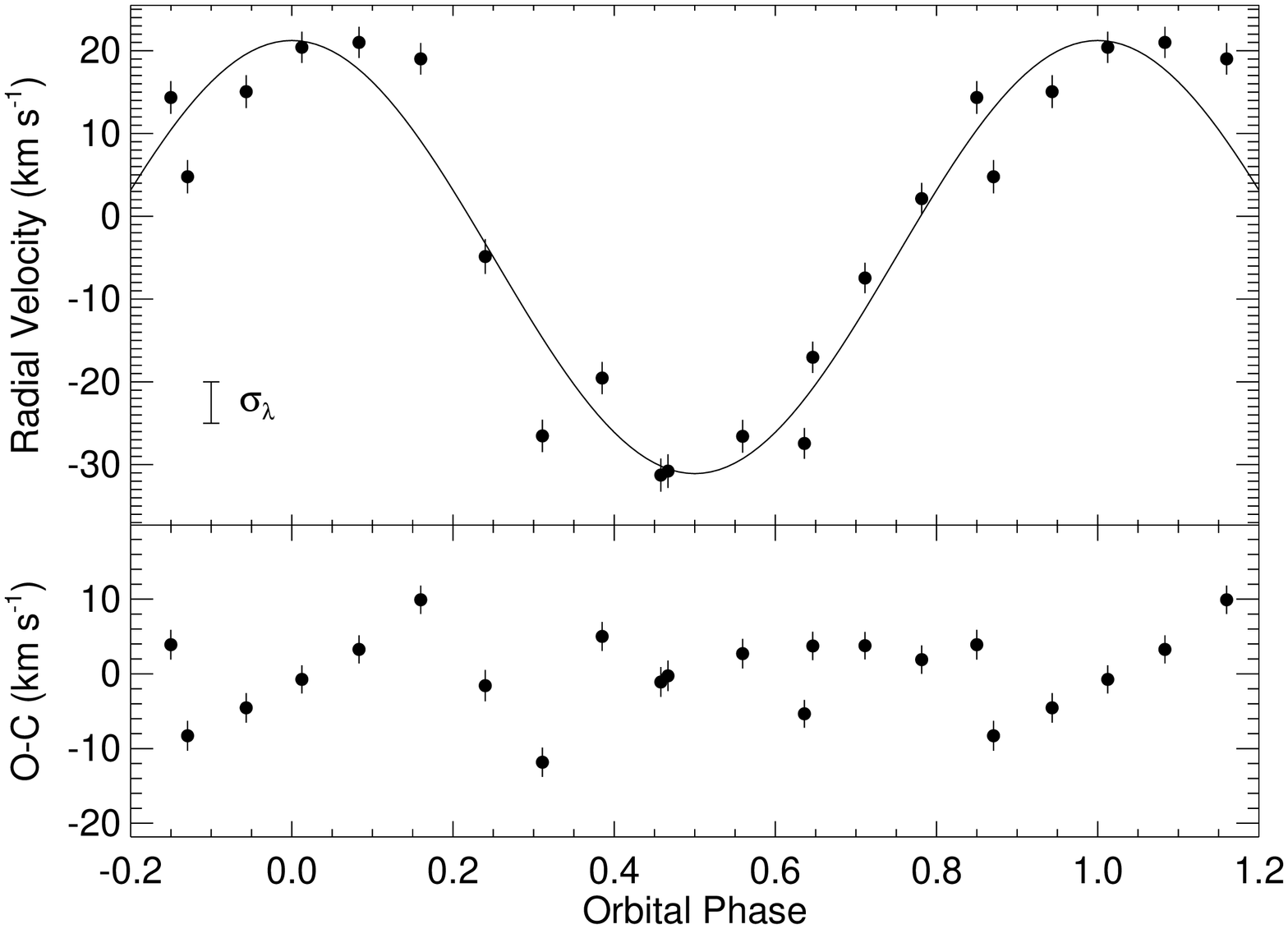}}
\end{center}
\caption{Radial velocities for HD 164536. Filled circles are data
  from this analysis, with uncertainties shown as line segments. 
  Plotted beneath the velocities are the observed minus calculated 
  values based on the circular orbital fit.}
\label{hd164536fit}
\end{figure}

\begin{figure}
\begin{center}
{\includegraphics[angle=90,height=12cm]{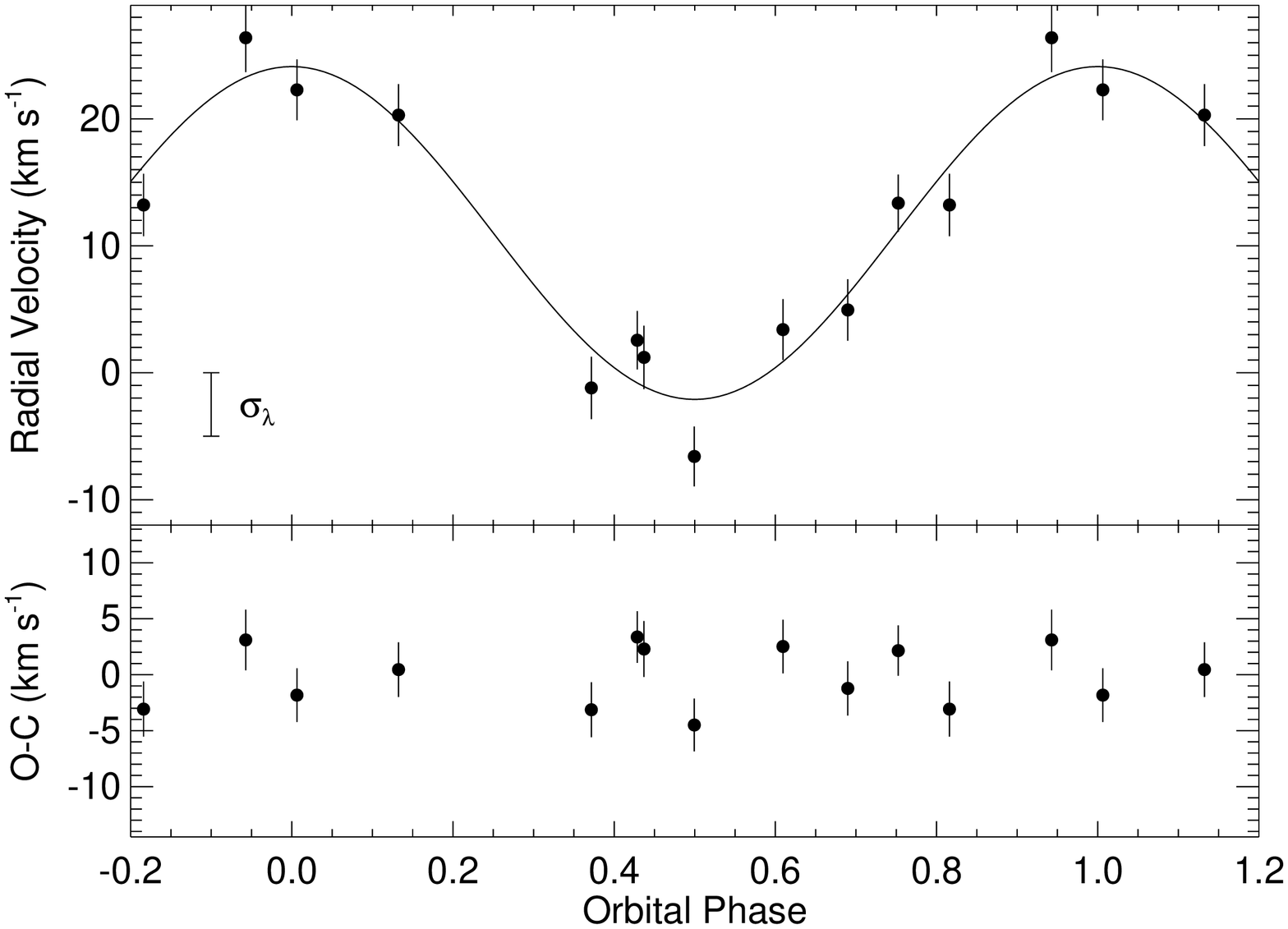}}
\end{center}
\caption{Radial velocities for BD$-16^\circ 4826$. Filled circles are
  data analyzed here, with uncertainties shown as line segments. 
  Plotted beneath the velocities are the observed minus calculated 
  values based on the circular orbital fit.}
\label{bd-164826fit}
\end{figure}

\begin{figure}
\begin{center}
{\includegraphics[angle=90,height=12cm]{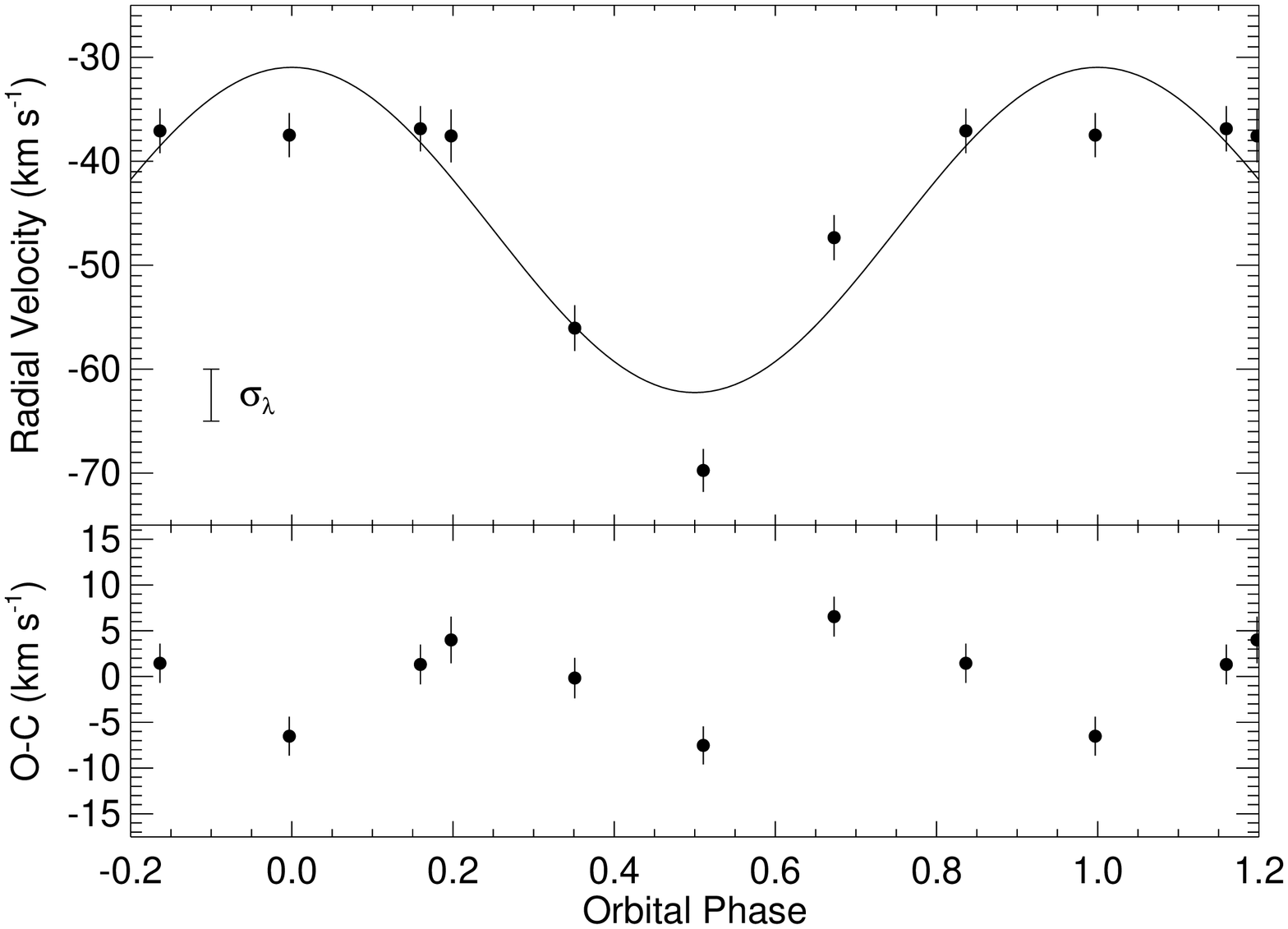}}
\end{center}
\caption{Radial velocities for HDE 229232. Filled circles are data
  from this analysis, with uncertainties shown as line segments. 
  Plotted beneath the velocities are the observed minus calculated 
  values based on the circular orbital fit.}
\label{hd229232fit}
\end{figure}

\begin{figure}
\begin{center}
{\includegraphics[angle=90,height=12cm]{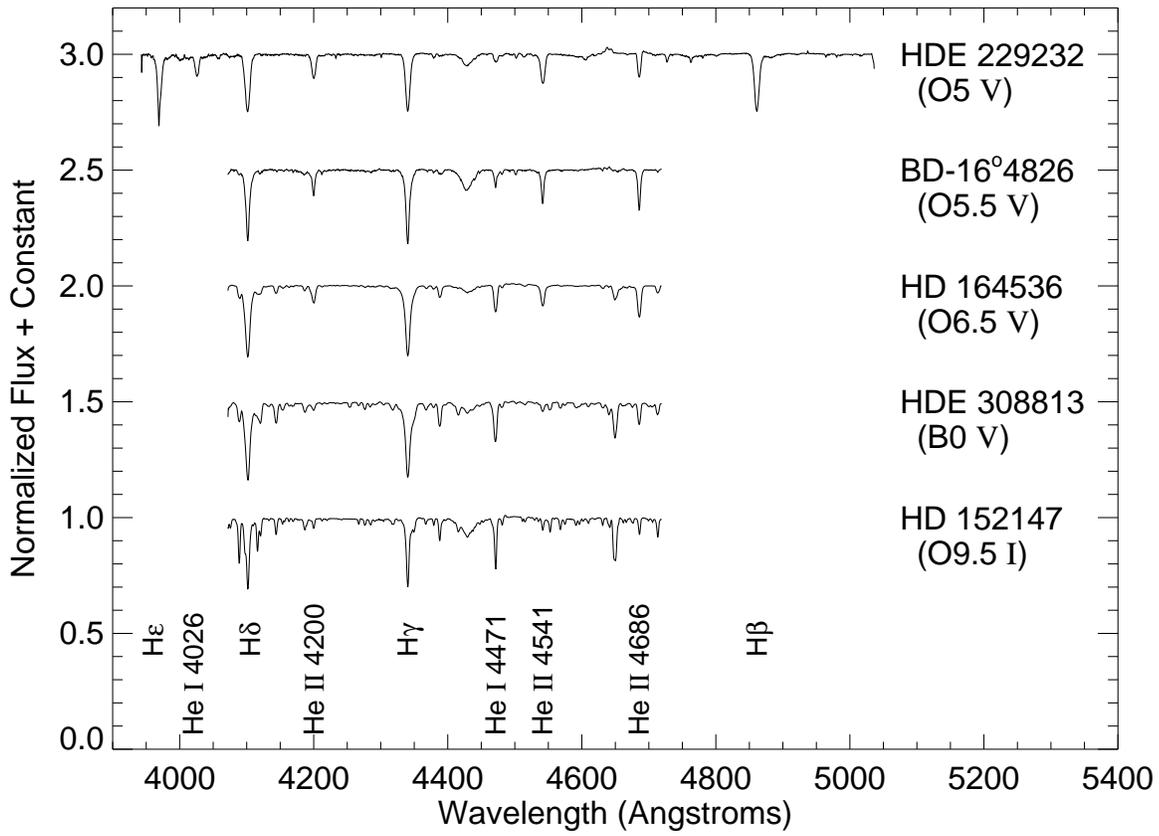}}
\end{center}
\caption{Spectra and prominent spectral features for objects
  discussed here. The hottest star is at top and targets are
  then plotted in order of decreasing effective temperature.}
\label{sb1combospec}
\end{figure}

For each object, we ran fits of the radial velocity 
measurements for eccentric orbits and
for circular orbits using the nonlinear, least-squares fitting
routine from \citet{mor74}. We note that we did not use
any other period search algorithm because our data sets 
covered all of or significant fractions of each orbital
period, so there was no ambiguity in the selection of an
initial period estimate for the general orbital fitting
scheme. Following \citet{luc71}, we computed a
statistic based on the rms of the fit for the eccentric and 
circular cases, $p=({\rm rms}_e/{\rm rms}_c)^{\beta}$, where 
${\rm rms}_e$ is the rms for the eccentric orbital fit, 
${\rm rms}_c$ that for the circular 
orbital fit, and $\beta=(N-M)$, where $N$ represents the number
of observations, and $M$ the number of parameters fit (6 for 
the eccentric case). Values of $p>0.05$ (the 5\% significance level) indicate
that the eccentric orbit does not statistically improve the fit. 
For each of the orbits computed here, the eccentric fits provided no
significant improvement in the residuals, so we adopted circular orbits. 
The circular orbital parameters of period, $P$, the time of 
maximum velocity, $T_{\rm 0}$, the systemic velocity for the system, 
$\gamma$, and the velocity semiamplitude, $K_1$,
are listed in Table~\ref{sb1orbitparams}.
Also listed are the mass function, $f(m)$, the projected orbital
semi-major axis, $a_1 \sin i$, the rms of the fit, and the number
of observations. The radial velocities and orbital solutions are plotted 
in Figures~\ref{hd308813fit} through \ref{hd229232fit}, as are the
uncertainties associated with the wavelength calibration, 
$\sigma_{\lambda}$. 
Using the final orbit fits we employed a  
shift-and-add algorithm to the spectrum stack in order 
to create a higher S/N master spectrum. 
These spectra are plotted in Figure~\ref{sb1combospec}.

Stellar parameters were estimated by fitting the master spectrum
for each object with model templates from OSTAR2002 \citep{lan03} 
and BSTAR2006 \citep{lan07}. Initial values for $v$ sin $i$ were
estimated by convolving the model template with the instrumental
broadening (see \S 2) and with a projected rotational velocity
($v$ sin $i$) function to obtain a broadening function making a rough
match of one prominent non-hydrogen absorption line. The choice of 
the absorption line used differed based on the temperature 
of the star. We adopted 
linear limb darkening coefficients from \citet{wad85}.
A least squares grid search
routine was then employed to refine the estimates of $T_{\rm eff}$
and log $g$ from fits to several spectral features (listed for each
individual object in the
following section). These updated estimates for $T_{\rm eff}$ and
log $g$ were then used to refine the $v$ sin $i$ value by
comparison of the FWHM measurements of lines in the master
spectrum with those from the model template
broadened for a range of $v$ sin $i$. The adopted
$T_{\rm eff}$ and $v$ sin $i$ values are
means of individual line measurements with uncertainties computed from
the standard deviation of the measurements of different lines. 
For log $g$, we
used a weighted average where Balmer line wing fits are
given twice as much weight in the calculation of means
because of their sensitivity to pressure broadening. 
It should be noted that we cannot measure $v~\rm{sin}$~$i$ values
less than roughly half our instrumental broadening, or about 50 km 
s$^{-1}$. Another limitation is that both OSTAR2002 
\citep{lan03} and BSTAR2006 \citep{lan07} are based on plane-parallel 
approximations and solar helium abundances. These approximations
may be less accurate for very high luminosity stars, such as HD 152147,
which generally have extended atmospheres and significant winds. 
Lower values for log $g$ indicate the star is evolved, 
and such stars may also have enhanced helium abundances due to mixing
and mass loss. The entire process of fitting stellar parameters 
was repeated once to ensure accuracy of the final results. 

The derived $T_{\rm eff}$, log $g$, and $v$ sin $i$ values are 
listed in Table~\ref{sb1params}, along with a spectral classification
estimate made by comparing our computed values for $T_{\rm eff}$
and log $g$ with Tables 4, 5, and 6 of \citet{mar05} for O-stars,
and Table 2 from \citet{boh81} for the two targets on the border
between O- and B-type stars, HDE 308813 and HD 152147. Column 2 in
the table lists the spectral classifications for the primary 
component found in the literature. The final three
columns in Table~\ref{sb1params} list the parameters used to obtain
a spectroscopic parallax distance estimate that is described in detail
for each object in the next section. 


\section{Discussion of Individual Objects}

\subsection{HDE 308813}

The variable velocity of HDE 308813 and its 
membership in cluster IC 2944 were first suggested by
\citet{tha65} based on three observations spaced two years apart.
They suggested that HDE 308813 may be an SB2 due to differences in 
velocities from \ion{He}{2} compared to other 
lines seen on their last plate.
\citet{hua06} also measured three velocities, over a time
span of 4 days, and suggested the system was a possible 
single-lined spectroscopic binary. \citet{san11}  
gathered 11 velocities over time spans of days to years, and they were 
therefore unable to discern an orbital period or ephemeris 
for this system, although they did confirm the SB1 nature
suggested by \citet{hua06}. The observations reported 
here have the benefit of being made over three separate epochs,
the second one a 16 day span. Our computed orbital 
period of 6.340 d (Table~\ref{sb1orbitparams}) 
is therefore covered twice in that time span. 
The radial velocity measurements from 
\citet{tha65} ($-$12, 18, and $-$40 km s$^{-1}$), 
\citet{ard77} ($-$10 km s$^{-1}$), and
\citet{hua06} (21.9,$-$13.4, and $-$17.6 km s$^{-1}$)
all span the range of velocities we observe. The data from
\citet{hua06} are overplotted in Figure~\ref{hd308813fit}, 
but were not used in computation of the orbit. Including any of the
archival data increased the rms of the fit by at least a factor of two
(presumably due to differences in the way the velocity was measured
by these authors), so we only fit data reported in this work.

Prior $v$~sin~$i$ measurements are 186$\pm$11 km s$^{-1}$ by 
\citet{daf07} (using \ion{He}{1} $\lambda\lambda4026, 4387, 4471$), 
196$\pm$9 km s$^{-1}$ by 
\citet{hua06} (who used the same lines plus \ion{Mg}{2} $\lambda4481$),
and 197 km s$^{-1}$ by \citet{pen04} (from UV lines
in spectra obtained by the  Hubble Space Telescope and Space
Telescope Imaging Spectrograph). 
All of these measurements agree within 1$\sigma$ of our
computed value of 204$\pm$10 km s$^{-1}$ obtained by measuring
\ion{He}{1} $\lambda4143, 4387, 4471$. $T_{\rm eff}$ and
log $g$ were estimated using H$\delta$, H$\gamma$, \ion{He}{1}
$\lambda\lambda4143,4387$, and \ion{He}{2} $\lambda4686$.
Based on the values computed above and the calibration of 
\citet{boh81}, the spectral classification of HDE 308813 is
B0~V. This is very close to the classification by \citet{sch70}
of O9.5~V. 

We can estimate the distance to HDE 308813 via spectroscopic 
parallax. \citet{tha65} measured $V=9.28$
and $(B-V)=0.04$ for HDE 308813.
We adopt $(B-V)_{\rm 0}=-0.26$ for a B0~V star from 
\citet{weg94} and $M_V=-3.90$ for an O9.5~V star from 
\citet{mar05}. Assuming $R_V=3.1$ leads
to a distance of 2.82 kpc, and this value is listed in Column
9 of Table~\ref{sb1params}. This distance puts HDE 308813 a
bit further than the distance for IC 2944 of
2.2 kpc from \citet{tom98}, but it may be a member of a
more distant association (see their Table 5). It should also 
be noted 
that the difference in the distances may be due to the uncertainties
in our estimation of $(B-V)$, $(B-V)_{\rm 0}$, $M_V$, and $R_V$.
However, \citet{dia02} report an average cluster velocity for IC 2944 of
$-$4.8$\pm$5.1 km s$^{-1}$. The systemic velocity from our orbital
fit for HDE 308813 is 8.5 km s$^{-1}$, 2.6$\sigma$ from the 
cluster average, making cluster membership unlikely.

\subsection{HD 152147}

HD~152147 lies in the vicinity of the Galactic open cluster
NGC 6231, but it is not likely to be a member. 
A number of velocities have been previously published,
five by \citet{str44}, one by 
\citet{con77b}, and 10 by \citet{lev88}, who were also 
the first to claim velocity variability. 
We fit an orbit to all these literature velocities and our 
own 15 measurements with equal weight and obtain the 
parameters listed in Table~\ref{sb1orbitparams} with the
orbit shown in Figure~\ref{hd152147fit}. 

\citet{con77a} measured $v$~sin~$i = 80$ km s$^{-1}$ 
by comparing H$\gamma$,
\ion{He}{1} $\lambda\lambda4387,4471$, and \ion{He}{2}
$\lambda4541$ with calibrated standard stars. 
We measure a rotational velocity of
150$\pm$28 km s$^{-1}$ from fits to \ion{Si}{4} $\lambda4088$,
\ion{He}{1} $\lambda\lambda4143,4471,4387$, and \ion{He}{2} $4686$.
Their spectra were obtained at higher resolution than our 
spectra, so our measurement should be taken as an upper 
limit. The estimates for $T_{\rm eff}$ and log $g$
were made from fits to H$\gamma$, \ion{He}{1} 
$\lambda\lambda4143,4387$, and \ion{He}{2} $\lambda4686$.
Using these numbers in concert with the calibration in 
\citet{mar05}, we estimate the spectral classification for
HD 152147 to be O9.5~I, very close to the spectral classification
given by \citet{wal72}, O9.7~Ib.

To estimate the spectroscopic parallax, we began with $V=7.23$ and
$(B-V)=0.37$ measured by \citet{sch69} from seven observations.  
\citet{weg94} gives a value of $(B-V)_{\rm 0}=-0.24$ for
an O9.5 I star. Further assuming a standard value of 
$R_V=3.1$ and using $M_V=-6.28$ in \citet{mar05} for
an O9.5 I star, leads to a calculated distance to HD~152147 of
2.11 kpc. This value does not agree well with the recent
distance estimate for HD~152147 of 1.64$\pm$0.24 kpc 
from \citet{mei09} that is based on interstellar line strength. 
Nor does this value match with the
distance estimate to NGC 6231 given by \citet{san06} of
1.64 kpc. This may mean our estimate for $R_V$ is
inaccurate for the line of sight to HD 152147, but this
also may be evidence suggesting that
HD 152147 is not a member of NGC 6231. For example,
\citet{hum78} places this star in the Sco OB1 association
at a distance of 1.91 kpc, a distance closer to our
estimate. Furthermore, \citet{kha05} reports that NGC 6231 has
an average cluster radial velocity of $-$27.28$\pm$2.98 
km s$^{-1}$, far different than our systemic velocity from
the orbital fit of $-$40.1$\pm$0.7 km s$^{-1}$.

\subsection{HD 164536}

The first radial velocity
measurement published is by \citet{hay32} who reports a
value of 6 km s$^{-1}$ from five spectral lines in a single
spectrum. Other radial velocities are averages from
several, typically unspecified, nights: 
$-$11.1 $\pm$ 3.2 km s$^{-1}$ from five plates 
reported by \citet{neu43} and $-$10 km s$^{-1}$ 
from three plates reported by \citet{fea55}. None 
of these velocities were used in the derivation of 
the orbital parameters listed in 
Table~\ref{sb1orbitparams}. 

HD~164536 is one of the brightest members of the cluster 
NGC~6530 \citep{wal57,bog89}. A proper motion study was made of 
NGC 6530 by \citet{van72}, who claim that HD~164536 is not a member.
However, its cluster membership was reestablished in
a series of papers based on \textit{IUE} spectra 
\citep{boh84,bog89,bog90}. \citet{afr75}
claimed detection of a nearby companion during a lunar occultation
observation at $\rho=0\farcs174\pm0\farcs003$ arcsec, 
$\theta=203^\circ$, and $\Delta m=2.2\pm0.4$ mag. Subsequent
follow up observations that had the ability to see such
a companion \citep{mas96,tur08} did not resolve this close 
companion. This star is the brightest
of a trapezium (4 star) system listed in the Washington Double 
Star Catalog (WDS; \citealt{mas01}). \citet{tur08} resolve 
the closest component at $1\farcs7$ distant, with other members outside
of their field of view. This close component is too faint 
($V=12.4$ according to the WDS) compared to HD 164536 
($V=7.1$; \citealt{hog00}) to influence our spectra.

To derive the effective temperature and log $g$,
H$\delta$, H$\gamma$, \ion{He}{1} $\lambda4471$, and 
\ion{He}{2} $\lambda\lambda4541,4686$ were used. The
projected rotational velocity was computed from 
observations of the \ion{He}{2} lines. \citet{bog89}
arrive at an effective temperature of 32 kK determined from
optical photometry and spectral
type of O9~V from their analysis of \textit{IUE} spectra. The value
derived here of 37.4 $\pm$ 0.9 kK is $>3\sigma$
higher, indicating an O6.5~V estimate of the spectral 
classification from Table 4 in \citet{mar05}. This classification
is also at odds with the optical classification reported by 
\citet{mac76} of O9~III, although the spectrum is noted as 
being overlapped with another star, making the classification
uncertain. \citet{hou88} find a spectral type of 
O7/O8, closer to our estimate. Given the uncertainties in
the spectral classifications in \citet{mac76} and 
\citet{bog89}, we think
our estimate of O6.5~V matches well with that of \citet{hou88},
and is currently the best estimate.

We estimate the distance to HD 164536 by taking the
Tycho $B$ and $V$ magnitudes \citep{hog00} transformed to Johnson 
magnitudes, an $M_V=-4.77$ for an 
O6.5~V \citep{mar05}, $(B-V)_0=-0.29$ \citep{weg94}, and
assume $R_V=3.1$. This results in a distance of 1.78 kpc,
equal to the 1.78 kpc estimate given by \citet{van72} and larger than the 
1.33 kpc distance from \citet{kha05}. Using $R_V=3.45$ for HD 164536,
as reported in \citet{weg03}, the distance drops to 1.71 kpc.
Both of the distances computed in this work agree within uncertainties
with the distance reported by \citet{van97} of 1.8$\pm$0.2 kpc. 
The systemic velocity
component of HD 164536 found here is $-$4.9$\pm$1.1 km s$^{-1}$; this
value is 1.3$\sigma$ from the cluster radial velocity of NGC 6530
listed by \citet{bar00} of $-$13.32$\pm$6.34 km s$^{-1}$. This appears
to offer tentative support for the inclusion of HD 164536 in NGC 6530.

\subsection{BD$-$16$^\circ$ 4826}

No earlier velocity measurements were found in the literature for 
BD$-16^\circ 4826$, making our 11 velocities and SB1 fit the 
first published. However, additional observations are desirable
because our observing span of 12 d is somewhat smaller than the
16 d orbital period. Consequently, the orbital parameters should
be considered as preliminary values.

BD$-16^\circ 4826$ is a member of the cluster NGC 6618 which is 
embedded in M 17. 
Estimates for effective temperature and log $g$ were made
from fits to H$\delta$, H$\gamma$, \ion{He}{2} 
$\lambda\lambda4200,4541,4686$ and \ion{He}{1} 
$\lambda\lambda4387,4471$. We used \ion{He}{2} 
$\lambda\lambda4541,4686$ 
to derive the $v$~sin~$i$ value listed in Table~\ref{sb1params}.
There was very large scatter in our attempts to fit the
spectrum of BD$-16^\circ 4826$, as evidenced by the
highest uncertainties for stellar parameters in this work.
Listed as an O5 in \citet{hil56}, the spectral classification
of this object has varied from O3~V based on optical spectra
\citep{gar91} to any of four spectral types ranging from 
O5~V to O9~III that were based upon IR data by \citet{pov09}. 
The best estimators we have available as luminosity criteria 
are the wings of the Balmer lines. These 
definitively rule out the possibility of BD$-16^\circ 4826$
being evolved (see Fig.~\ref{sb1combospec} and 
Table~\ref{sb1params}), and we
estimate the spectral classification for BD$-16^\circ 4826$
to be O5.5~V. 

The
most recent optical photometry is reported by \citet{ogu76}.
They find, from two observations, $V=9.85$ and $(B-V)=0.94$.
Using $M_V=-5.07$ \citep{mar05}, $(B-V)_{\rm 0}=-0.30$ 
\citep{weg94}, and a standard value
$R_V=3.1$, we arrive at a spectroscopic distance to
BD$-16^\circ 4826$ of 1.60 kpc. The distance to NGC 6618
and M 17 is a topic of some debate. The recent IR study
by \citet{pov09} lists two distances, $1.6\pm0.3$ 
kpc from \citet{nie01} and $2.1\pm0.2$ kpc from \citet{hof08}. 
The value derived here agrees with the \citet{nie01} estimate. 
Our computed value is also within the uncertainty of the lower 
limit on the
distance reported by \citet{han97} of $1.3^{+0.4}_{-0.2}$ kpc
based on extinction corrected IR magnitudes of O-stars. The 
systemic velocity found here for BD$-16^\circ 4826$ of 
$11.0\pm1.0$ km s$^{-1}$ is 2.2$\sigma$ different than
the cluster velocity of $-17\pm13$ km s$^{-1}$ 
\citep{kha05}. This discrepancy in velocities hints that it
is less likely that BD$-16^\circ 4826$ is actually a 
member of NGC 6618.

\subsection{HDE 229232}

There is very little in the literature about HDE 229232.
Like with BD$-16^\circ4826$, our seven velocities are the first
published. However, with only seven data points, 
our orbit should be considered preliminary. For example,
it is possible that the system has a significant eccentricity
and longer period, so that our data record the variations around
a single periastron point. Our estimate that the star is an
O5~V with a variable velocity indicates that this is a 
massive binary that deserves additional observational work.

To estimate effective temperature and log $g$, we used the
H$\delta$ wings, \ion{He}{1} $\lambda4026$, and \ion{He}{2}
$\lambda\lambda4200,4541$. Only the helium lines were used
to compute the $v$~sin~$i$. Our estimated spectral classification
of O5~V is very close to that listed by \citet{wal73} of 
O4~Vn((f)).

Using $V=9.525$ and $(B-V)=0.82$ from \citet{hil56}, 
$M_V=-5.21$ from \citet{mar05} and $(B-V)_o=-0.30$ 
\citep{weg94} (both values
for an O5~V star), and the standard $R_V=3.1$, the distance 
to HDE 229232 is 1.79 kpc. The galactic coordinates of HDE 229232
are $\ell=77\fdg40$ and $b=0\fdg93$, looking down the Cygnus
arm, where many associations with a range in distances from
0.75 to 2.32 kpc appear more or less in the same part of the sky 
\citep[][and references therein]{uya01}. It lies in the direction
of the Cyg OB8 association that \citet{hum78} places
at a distance of 2.29 kpc. The systemic velocity from the 
orbital fit of HDE 229232 is $-$46.6$\pm$0.8 km s$^{-1}$.
\citet{mel09} report a velocity for Cyg OB8 from velocities
in the work of \citet{bar00} of $-$21$\pm$11 km s$^{-1}$.
This puts our estimate of the systemic velocity 2.3$\sigma$
from the mean value for Cyg OB8, perhaps lending support
that HDE 229232 is not a member of the Cyg OB8 association,
but possibly a member of one of the other associations in
that direction.


\section{Discussion}

The five systems studied here are all relatively long period, low
semiamplitude SB1s. The velocity semiamplitudes of these systems
range from 13.1 km s$^{-1}$ to 26.2 km s$^{-1}$ with orbital
periods ranging from 6.2 to 15.8 days. 
None of these systems show eclipses or
photometric variation in either the All Sky Automated Survey
\citep{poj02} nor The Amateur Sky Survey
(TASS\footnote{http://sallman.tass-survey.org/servlet/markiv/}; 
\citealt{dro06}). The only object with published time
series photometry is HD 152147, and \citet{ste93} found no variability.
BD$-16^\circ 4826$ is an X-ray source but
with a flux typical of single O-stars \citep{naz09}. 

For circular orbits, the orbital semiamplitude is given by
$$K_1 = 289~{\rm km~s}^{-1} (M_1/ 25 M_\odot)^{1/3} 
(P/ 10~{\rm d})^{-1/3} {q \over {(1+q)^{2/3}}} \sin i$$
where the mass ratio is $q=M_2/M_1$.
Therefore the  low $K_1$ systems in our sample are drawn 
from those with low mass companions (small $q$) and/or low
inclination. 
We can estimate the most probable $q$ from the statistical
method of \citet{maz92}. This method requires assumptions
about the mass ratio distribution of the sample and
the mass of the primary, which we estimate from \citet{mar05}. 
Based on observations of O-stars \citep{mas98}, we
assume the mass ratio distribution is uniform.
The resultant mass ratios ($q$) from this distribution 
are listed in Table~\ref{sb1comps}. Estimates for
the spectral classification of the 
secondary are made by comparing the mass in column 5
to the values listed in Table B.1 of \citet{gra05}. 
We assume these companions are all main sequence stars,
because all the primary stars are so young. 
Column 7 lists the difference in $V$ magnitude between
the primary and secondary, again as estimated from 
\citet{gra05}. These are all greater than
2.3 mag, and they lead to the monochromatic flux ratios 
given in column 8. The extremely small flux ratios may be 
part of the reason why companions are not seen in the data. 

The true nature of the mass ratio distribution for 
O-type stars is unknown. (For a discussion of the current
state of our understanding of the mass ratio distribution,
see the recent work by \citet{kim12}.)
To explore the range of
possibilities besides the flat distribution used above,
we take two extreme values for a power law distribution
$N(q)=N_0~q^{\alpha}$. The first is the standard
Salpeter distribution \citep{sal55} with an exponent of 
$\alpha = -1.35$. The second has an exponent of
$\alpha = 1.0$, a distribution that is more biased toward
near equal-mass binaries, as was suggested in \citet{pin06}. 
With these mass ratio distributions, and estimates
for the masses of the primary and secondary,
we can solve for the probable inclination of the orbit based
on the mass function. The resulting ranges of mass ratio
and inclination are listed in Table~\ref{qitab}. These
ranges are not large, and the estimates for 
orbital inclination are low ($i<37^{\circ}$).

Taking our values for $i$ and orbital period, combined with
an estimate of the stellar radius \citep{mar05}, we
compute the synchronous rotational velocity, 
$v \sin i$~(sync), listed in column 10 of 
Table~\ref{sb1comps}. These numbers are all much
smaller than the $v \sin i$ measurements listed in 
column 5 of Table~\ref{sb1params}, which is consistent with
the idea that these systems are too young to have attained 
synchronous rotation. 

Do the line blending effects of the companions influence 
our results for $v \sin i$ or
$K_1$? To check the magnitude of the effect, we can consider
the maximum separation of the spectral features of the 
individual components. The ratio of maximum Doppler
separation to projected rotational velocity is
$${{K_1(1+1/q)} \over {v \sin i}}.$$
By using the mass ratio as computed above, we calculate
this separation and list it in column 9 of 
Table~\ref{sb1comps}. These values are all well below unity,
meaning the lines of the components are always blended with
those of the primary (except in cases where a lower ionization
line appears in the secondary but not in the primary spectrum). 
The additional fact that
the companions are extremely faint means that line
blending problems related to measurements of 
$v \sin i$ and $K_1$ will be minimal. As an example, 
we introduced a model spectrum corresponding
to a B3~V star, as estimated for the companion to 
BD$-16^\circ 4826$ (Column 6, Table~\ref{sb1comps}), 
at a shift of 
$K_1(1+1/q)$ and normalized according to the flux ratio. 
This object was chosen because it has the largest ratio
of $K_1(1+1/q)/v \sin i$, so it should show the largest
affect of blending problems. 
We chose a $v \sin i$ of the companion of 20 km s$^{-1}$.
Such very narrow lines would have more influence
on the measured parameters of the primary than wider 
shallower lines.
Measuring the velocity in the same way as described
above, we compute a maximum offset of 3.6 km s$^{-1}$ for 
velocity measurements, comparable to the measurement 
uncertainties on an individual spectrum. 
We similarly added a secondary model spectrum at 
$\pm K_1 (1+1/q)$ to each of the lines used to 
measure $v \sin i$, and calculated a value 5.0 km s$^{-1}$
higher than previously. Therefore, even for the narrowest 
lined star in our sample, the uncertainty in $v \sin i$
introduced by the companion is on the order of 5 
km s$^{-1}$. Although the effects are small, the influence
of a companion spectrum on the velocity measurements
will be to pull the measurement back toward the rest
wavelength of the line. This will make $K_1$ lower, the
mass function lower, and thus the estimate of mass ratio and 
orbital inclination lower. Consequently, the values of 
$q$ and $i$ in Table~\ref{qitab} are therefore probably 
lower than the true values.

Companions will also have an influence on $T_{\rm eff}$
and log $g$ estimates. In the spectral range covered
by our data, an O-star has several \ion{He}{2} lines that
a mid B-star does not. Therefore, the effect will be to
make the O-star lines appear shallower, making the
O-star appear cooler. \ion{He}{1} lines are stronger
in B-stars, and in combination with an O-star spectrum will make
these lines appear deeper, again causing the O-star spectrum to
appear cooler. Thus, our estimates for $T_{\rm eff}$
and log $g$ are also lower limits and are likely more
indicative of the combined spectrum of the binary.

Our results from this analysis suggest that all the objects
in this sample are SB1. This is due to three main effects: 
low orbital inclination, broad spectral features of 
the primary stars, and faint companions. 
In the future, high resolution and high S/N spectra may be 
able to extract the companion spectrum in systems like 
those studied here \citep{gie94}.
The next paper in this series will describe 
double-lined spectroscopic orbits for Galactic O-type stars
where companion spectra are clearly seen. 


\acknowledgments

We are grateful to the directors and staffs of KPNO and CTIO for their
help in obtaining these observations. This material is based upon work 
supported by the National Science Foundation under 
Grant No.\ AST-0606861 and AST-1009080. This research has made use
of the Washington Double Star Catalog maintained at the U.S. Naval
Observatory.

{\it Facility:} \facility{CTIO:1.5m}
{\it Facility:} \facility{KPNO:2.1m}




\newpage

\begin{deluxetable}{cccrrr}
\tabletypesize{\scriptsize}
\tablewidth{0pt}

\tablecaption{Radial Velocity Measurements\label{sb1rvs}}
\tablehead{
  \colhead{Star}               &
  \colhead{Date}               &
  \colhead{Orbital}            &
  \colhead{$V_r$}              &
  \colhead{$\sigma$}           &
  \colhead{$O-C$}              \\
  \colhead{Name}               &
  \colhead{(HJD -- 2,450,000)} &
  \colhead{Phase}              &
  \colhead{(km s$^{-1}$)}      &
  \colhead{(km s$^{-1}$)}      &
  \colhead{(km s$^{-1}$)}      }
\startdata
HDE 308813 & 2714.805 & 0.699 & --0.6  & 1.7 &  --2.2 \\
HDE 308813 & 2715.708 & 0.841 & 20.6   & 1.7 &    4.6 \\
HDE 308813 & 2715.840 & 0.862 & 31.2   & 1.8 &    8.2 \\
HDE 308813 & 2716.707 & 0.999 & 27.3   & 1.7 &  --3.3 \\
HDE 308813 & 2716.866 & 0.024 & 20.7   & 1.9 &  --9.7 \\
HDE 308813 & 2717.697 & 0.155 & 32.8   & 1.9 &   12.0 \\
HDE 308813 & 2717.801 & 0.171 & 16.3   & 1.8 &  --2.6 \\
HDE 308813 & 2718.699 & 0.313 & 4.5    & 1.8 &    4.7 \\
HDE 308813 & 3017.743 & 0.484 & --13.6 & 2.0 &    0.0 \\
HDE 308813 & 3018.753 & 0.644 & --1.6  & 1.8 &    3.6 \\
HDE 308813 & 3019.735 & 0.799 & 16.2   & 2.0 &    1.0 \\
HDE 308813 & 3019.836 & 0.815 & 26.0   & 1.9 &    8.8 \\
HDE 308813 & 3020.718 & 0.954 & 23.6   & 1.9 &  --6.1 \\
HDE 308813 & 3021.713 & 0.111 & 28.1   & 1.9 &    2.6 \\
HDE 308813 & 3021.816 & 0.127 & 29.6   & 1.9 &    5.7 \\
HDE 308813 & 3022.721 & 0.270 & --3.7  & 1.9 &  --9.5 \\
HDE 308813 & 3022.865 & 0.292 & --4.7  & 1.9 &  --7.4 \\
HDE 308813 & 3023.702 & 0.424 & --18.5 & 2.0 &  --7.3 \\
HDE 308813 & 3023.806 & 0.441 & --11.2 & 1.9 &    0.9 \\
HDE 308813 & 3027.727 & 0.059 & 24.9   & 2.0 &  --4.2 \\
HDE 308813 & 3028.704 & 0.213 & 16.4   & 2.0 &    3.0 \\
HDE 308813 & 3028.787 & 0.227 & 19.6   & 2.0 &    7.8 \\
HDE 308813 & 3029.744 & 0.377 & --3.8  & 2.0 &    3.7 \\
HDE 308813 & 3030.694 & 0.527 & --11.2 & 1.9 &    2.1 \\
HDE 308813 & 3030.776 & 0.540 & --7.9  & 2.0 &    5.1 \\
HDE 308813 & 3031.712 & 0.688 & --1.4  & 1.9 &  --1.5 \\
HDE 308813 & 3032.636 & 0.834 & 10.0   & 1.9 &  --9.6 \\
HDE 308813 & 3032.782 & 0.857 & 24.5   & 2.0 &    2.2 \\
HDE 308813 & 3033.713 & 0.004 & 34.9   & 1.9 &    4.3 \\
HDE 308813 & 3159.469 & 0.840 & 15.0   & 1.8 &  --5.4 \\
HDE 308813 & 3160.463 & 0.997 & 26.8   & 1.8 &  --3.8 \\
HDE 308813 & 3162.531 & 0.323 & --1.9  & 1.8 &  --0.5 \\
HDE 308813 & 3163.532 & 0.481 & --20.8 & 1.9 &  --7.3 \\
\hline
HD 152147 & 3152.593 & 0.668 & --46.7 & 2.6 &  --0.2 \\
HD 152147 & 3153.620 & 0.742 & --33.6 & 2.5 &    7.2 \\
HD 152147 & 3154.580 & 0.811 & --43.7 & 2.5 &  --8.6 \\
HD 152147 & 3154.811 & 0.828 & --35.1 & 2.5 &  --1.2 \\
HD 152147 & 3155.743 & 0.896 & --32.0 & 2.7 &  --2.3 \\
HD 152147 & 3156.775 & 0.970 & --23.8 & 3.0 &    3.5 \\
HD 152147 & 3157.764 & 0.042 & --26.7 & 2.5 &    0.7 \\
HD 152147 & 3157.874 & 0.050 & --40.4 & 3.4 & --12.7 \\
HD 152147 & 3158.696 & 0.109 & --29.2 & 2.5 &    0.8 \\
HD 152147 & 3159.714 & 0.183 & --35.0 & 2.4 &  --0.2 \\
HD 152147 & 3160.721 & 0.256 & --41.6 & 2.5 &  --1.0 \\
HD 152147 & 3161.737 & 0.329 & --39.2 & 2.5 &    7.1 \\
HD 152147 & 3162.688 & 0.398 & --60.2 & 2.6 &  --9.5 \\
HD 152147 & 3163.709 & 0.472 & --50.8 & 2.6 &    2.2 \\
HD 152147 & 3164.599 & 0.536 & --57.7 & 2.7 &  --4.8 \\
\hline
HD 164536 & 3152.753 & 0.559 & --26.6 & 2.0 &   2.7  \\
HD 164536 & 3153.777 & 0.636 & --27.4 & 1.9 & --5.3  \\
HD 164536 & 3153.916 & 0.646 & --17.0 & 1.9 &   3.7  \\
HD 164536 & 3154.781 & 0.711 &  --7.5 & 1.8 &   3.8  \\
HD 164536 & 3155.720 & 0.782 &    2.1 & 1.9 &   1.9  \\
HD 164536 & 3156.633 & 0.850 &   14.4 & 2.0 &   3.9  \\
HD 164536 & 3156.910 & 0.871 &    4.8 & 2.0 & --8.3  \\
HD 164536 & 3157.881 & 0.943 &   15.1 & 2.0 & --4.5  \\
HD 164536 & 3158.803 & 0.012 &   20.4 & 1.9 & --0.7  \\
HD 164536 & 3159.750 & 0.083 &   21.0 & 1.9 &   3.2  \\
HD 164536 & 3160.773 & 0.160 &   19.0 & 1.9 &   9.9  \\
HD 164536 & 3161.842 & 0.240 &  --4.9 & 2.1 & --1.5  \\
HD 164536 & 3162.789 & 0.311 & --26.5 & 2.0 & --11.8 \\
HD 164536 & 3163.779 & 0.385 & --19.5 & 1.9 &   5.0  \\
HD 164536 & 3164.751 & 0.458 & --31.3 & 2.0 & --1.1  \\
HD 164536 & 3164.870 & 0.467 & --30.8 & 2.0 & --0.3  \\
\hline
BD$-16^\circ 4826$ & 3152.899 & 0.391 & --0.7 & 2.5 & --0.5 \\
BD$-16^\circ 4826$ & 3153.797 & 0.445 & 2.9   & 2.3 &   5.4 \\
BD$-16^\circ 4826$ & 3153.929 & 0.453 & 1.6   & 2.5 &   4.3 \\
BD$-16^\circ 4826$ & 3154.913 & 0.513 & --6.1 & 2.3 & --2.8 \\
BD$-16^\circ 4826$ & 3156.646 & 0.617 & 3.9   & 2.4 &   3.5 \\
BD$-16^\circ 4826$ & 3157.914 & 0.693 & 5.2   & 2.4 & --0.7 \\
BD$-16^\circ 4826$ & 3158.901 & 0.753 & 13.8  & 2.3 &   2.6 \\
BD$-16^\circ 4826$ & 3159.902 & 0.813 & 13.4  & 2.5 & --2.9 \\
BD$-16^\circ 4826$ & 3161.900 & 0.934 & 26.6  & 2.7 &   2.9 \\
BD$-16^\circ 4826$ & 3162.901 & 0.994 & 22.3  & 2.4 & --2.6 \\
BD$-16^\circ 4826$ & 3164.888 & 0.114 & 20.2  & 2.4 & --1.2 \\
\hline
HDE 229232 & 4785.649 & 0.198 &  --37.5 & 2.5 & 4.0   \\
HDE 229232 & 4786.594 & 0.351 &  --56.1 & 2.2 & --0.2 \\
HDE 229232 & 4787.576 & 0.511 &  --69.7 & 2.1 & --7.5 \\
HDE 229232 & 4788.576 & 0.673 &  --47.4 & 2.2 & 6.5   \\
HDE 229232 & 4789.583 & 0.836 &  --37.1 & 2.1 & 1.4   \\
HDE 229232 & 4790.572 & 0.997 &  --37.5 & 2.1 & --6.5 \\
HDE 229232 & 4791.574 & 0.160 &  --36.9 & 2.1 & 1.3   \\
\enddata
\end{deluxetable}

\clearpage

\begin{deluxetable}{lccccc}
\tabletypesize{\scriptsize}
\tablewidth{0pt}

\tablecaption{Circular Orbital Elements\label{sb1orbitparams}}
\tablehead{
\colhead{Parameter}          &
\colhead{HDE 308813}          &
\colhead{HD 152147}          &
\colhead{HD 164536}          &
\colhead{BD$-16^\circ 4826$} &
\colhead{HDE 229232}          }
\startdata
$P(\rm d)$                           & 6.340$\pm$0.004  & 13.8194$\pm$0.0003 & 13.4$\pm$0.6    & 15.8$\pm$1.3        & 6.2$\pm$0.2     \\
$T_{\rm 0} (\rm{HJD} - 2,400,000.0)$ & 53033.69$\pm$0.08& 31212.0$\pm$0.2    & 53158.6$\pm$0.8 & 53162.8$\pm$1.8     & 54790.6$\pm$0.1 \\
$\gamma$ (km s$^{-1}$)               & 8.5$\pm$1.1      & --40.1$\pm$0.7     & --4.9$\pm$1.1   & 11.0$\pm$1.0        & --46.6$\pm$0.8  \\
$K_1$ (km s$^{-1}$)                  & 22.2$\pm$1.5     & 13.1$\pm$1.0       & 26.2$\pm$1.5    & 13.1$\pm$1.6        & 15.6$\pm$1.2    \\
$f(m)$ ($M_{\rm \odot}$)             & 0.007$\pm$0.001  & 0.0032$\pm$0.0007  & 0.025$\pm$0.005 & 0.004$\pm$0.002     & 0.002$\pm$0.001 \\
$a_1$~sin~$i$ ($R_{\odot}$)          & 2.8$\pm$0.2      & 3.6$\pm$0.3        & 6.9$\pm$0.5     & 4.1$\pm$0.6         & 1.9$\pm$0.2     \\
rms (km s$^{-1}$)                    & 6.1              & 5.7                & 6.1             & 3.2                 & 7.3             \\
$N$                                  & 33               & 31                 & 16              & 11                  & 7               \\
\enddata
\end{deluxetable}

\begin{deluxetable}{lcccccccc}
\tabletypesize{\scriptsize}
\tablewidth{0pt}

\tablecaption{Individual Stellar Parameters\label{sb1params}}
\tablehead{
\colhead{}                        &
\colhead{Spectral}                &
\colhead{$T_{\rm eff}$}           &
\colhead{log $g$}                 &
\colhead{$v$ sin $i$}             &
\colhead{Spectral Classification} &
\colhead{$M_V$}                   &
\colhead{$E(B-V)$}                &
\colhead{Distance}                \\
\colhead{Object Name}             &
\colhead{Classification}          &
\colhead{(kK)}                    &
\colhead{(cm s$^{-2}$)}           &
\colhead{(km s$^{-1}$)}           &
\colhead{from $(T_{\rm eff}$,log $g)$} &
\colhead{(mag)}                   &
\colhead{(mag)}                   &
\colhead{(kpc)}                   \\
\colhead{(1)}                     &
\colhead{(2)}                     &
\colhead{(3)}                     &
\colhead{(4)}                     &
\colhead{(5)}                     &
\colhead{(6)}                     &
\colhead{(7)}                     &
\colhead{(8)}                     &
\colhead{(9)}                     }
\startdata
HDE 308813         & O9.5~V     & 29.9$\pm$0.3 & 3.73$\pm$0.09 & 204$\pm$10 & B0~V\tablenotemark{a}   & --3.90 & 0.30 & 2.82 \\
HD 152147          & O9.7~Ib    & 27.8$\pm$0.5 & 3.10$\pm$0.06 & 150$\pm$28 & O9.5~I\tablenotemark{a} & --6.28 & 0.61 & 2.11 \\
HD 164536          & O9~III     & 37.4$\pm$0.9 & 4.25$\pm$0.17 & 230$\pm$14 & O6.5~V\tablenotemark{b} & --4.77 & 0.22 & 1.78 \\
BD$-16^\circ 4826$ & O5         & 39.9$\pm$6.3 & 4.04$\pm$0.40 & 131$\pm$28 & O5.5~V\tablenotemark{b} & --5.07 & 1.24 & 1.60 \\
HDE 229232         & O4~Vn((f)) & 41.7$\pm$1.3 & 4.05$\pm$0.14 & 273$\pm$19 & O5~V\tablenotemark{b}   & --5.21 & 1.12 & 1.79 \\
\enddata
\tablenotetext{a}{Calibration of \citet{boh81}.}
\tablenotetext{b}{Calibration of \citet{mar05}.}
\tablecomments{Spectral classification references: HDE~308813: \citet{sch70}; 
  HD~152147: \citet{wal72}; HD~164536: \citet{mac76}; BD$-16^\circ 4826$: \citet{hil56}; 
  HDE~229232: \citet{wal73}.}
\end{deluxetable}
\clearpage

\begin{deluxetable}{llcccccccc}
\tabletypesize{\scriptsize}
\tablewidth{0pt}

\tablecaption{Probable Unseen Companion Parameters\label{sb1comps}}
\tablehead{
\colhead{}                         &
\colhead{$f(m)$}                   &
\colhead{$M_1$}                    &
\colhead{}                         &
\colhead{$M_2$}                    &
\colhead{Estimated Secondary}      &
\colhead{$\Delta V$}               &
\colhead{Flux Ratio}               &
\colhead{\underline{$K_1(1+1/q)$}} &
\colhead{$v \sin i$~(sync)}        \\
\colhead{System}                   &
\colhead{($M_{\odot}$)}            &
\colhead{($M_{\odot}$)}            &
\colhead{$q$}                      &
\colhead{($M_{\odot}$)}            &
\colhead{Spectral Classification}  &
\colhead{(mag)}                    &
\colhead{($f_2/f_1$)}              &
\colhead{$v \sin i$}               &
\colhead{(km s$^{-1}$)}            \\
\colhead{(1)}                      &
\colhead{(2)}                      &
\colhead{(3)}                      &
\colhead{(4)}                      &
\colhead{(5)}                      &
\colhead{(6)}                      &
\colhead{(7)}                      &
\colhead{(8)}                      &
\colhead{(9)}                      &
\colhead{(10)}                     }
\startdata
HDE 308813          & 0.007 & 14 & 0.26 & 3.4 & B7~V & 2.7 & 0.08 & 0.53 & 20 \\
HD 152147	    & 0.003 & 28 & 0.18 & 5.0 & B4~V & 4.8 & 0.01 & 0.55 & 23 \\
HD 164536 	    & 0.025 & 28 & 0.30 & 8.3 & B2~V & 2.3 & 0.12 & 0.49 & 13 \\
BD$-16^\circ 4826$  & 0.004 & 34 & 0.19 & 6.4 & B3~V & 3.1 & 0.06 & 0.67 & \phn9 \\
HDE 229232          & 0.002 & 38 & 0.16 & 5.9 & B3~V & 3.2 & 0.05 & 0.45 & 29 \\
\enddata
\end{deluxetable}

\begin{deluxetable}{lccccccc}
\tabletypesize{\scriptsize}
\tablewidth{0pt}

\tablecaption{Mass Ratios and Inclinations for Various Mass Ratio Distributions\label{qitab}}
\tablehead{
\colhead{}                        &
\multicolumn{3}{c}{$q$}           &&
\multicolumn{3}{c}{$i(^{\circ})$} \\
\noalign{\smallskip}
\cline{2-4}
\cline{6-8}
\noalign{\smallskip}
\colhead{System}         &
\colhead{$\alpha=-1.35$} &
\colhead{$\alpha=0.0$}   &
\colhead{$\alpha=1.0$}   &&
\colhead{$\alpha=-1.35$} &
\colhead{$\alpha=0.0$}   &
\colhead{$\alpha=1.0$}   }
\startdata
HDE 308813        &  0.15 & 0.26 & 0.43 && 35.5 & 20.5 & \phn6.0 \\
HD 152147         &  0.10 & 0.19 & 0.38 && 30.4 & 16.6 &    10.5 \\
HD 164536         &  0.18 & 0.30 & 0.45 && 37.0 & 22.7 &    12.6 \\
BD$-16^\circ4826$ &  0.11 & 0.19 & 0.39 && 32.4 & 18.3 &    13.1 \\
HDE 229232        &  0.09 & 0.16 & 0.36 && 28.0 & 15.4 & \phn7.4 \\
\enddata
\end{deluxetable}


\clearpage

%
%

\newpage

\newpage

\newpage

\newpage

\newpage

\input{epsf}

\end{document}